# FinFET and Nanowire SRAM Radiation Hardness Studies using *Ab initio*-TCAD Simulation Framework

Johan Saltin, Adam Elwailly, and Hiu Yung Wong[*]
Electrical Engineering Department
San Jose State University
San Jose, CA, [*]hiuyung.wong@sjsu.edu

*Abstract*—In this paper, we study the vulnerability of 5nm node FinFET and nanowire (20nm gate length) and their corresponding SRAM under radiation. *Ab initio* tools, SRIM, PHITS, and GEANT4, are used to find the Linear Energy Transfer (LET) of neutron and alpha particles in Silicon and Silicon-Germanium. Technology Computer-Aided Design (TCAD) is then used to find the most vulnerable incident location and direction in FinFET, nanowire, and their SRAM. Full 3D TCAD simulation, which allows the study of layout effect in SRAM, is used. It is found that NW is about 2-3 times more robust than FinFET in terms of flipping energy. Based on the simulation in the *ab initio*-TCAD framework, it is projected that there is a possibility to design an SRAM using NW that is immune to α-particle. It is also expected that SRAM can be optimized for more robust radiation hardness if Design Technology Co-Optimization (DTCO) is taken into consideration.

## I. INTRODUCTION

To control the short channel effect and enhance transistor performance, FinFET has already been used in the advanced technology nodes. It is expected in "2.1nm" node, nanowire (NW) or nanosheet are needed to enable further scaling [1]. In advanced technologies, design technology co-optimization (DTCO) becomes more important [2]. However, radiation hardness is usually not a part of the DTCO metrics.

SRAM radiation hardness is very important in mission-critical applications. Although there are many studies of SRAM radiation hardness in TCAD, most of them are mixed-mode simulations and do not consider layout effects [3][4] and, thus, cannot be used with DTCO. In this paper, an *ab initio*-TCAD framework is established (Fig. 1). We first use *ab initio* tools, GEANT4 [5], SRIM [6], and PHITS [7], to correlate alpha (α−) particle and neutron energies to Linear Energy Transfer (LET) in SiliconGermanium ($Si_{1-x}Ge_x$) with various mole fractions, *x*. Then, as an example, TCAD simulations are performed on Si FinFET and NW with $L_G$=20nm (5nm technology node) to find their radiation hardness in terms of the most vulnerable incident locations and directions. Finally, their SRAM's radiation hardness is simulated using full 3D TCAD simulations in which the layout effect can be taken into account. The result is then correlated with the most vulnerable α−particle and neutron energies. _Such an ab initio-TCAD approach allows the radiation hardness study of novel materials and technologies on any standard cells._

Corresponding author: Hiu Yung Wong

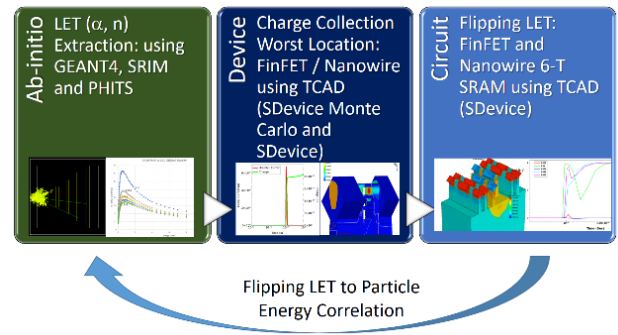

Fig. 1. *Ab initio*-TCAD simulation framsework used in this study.

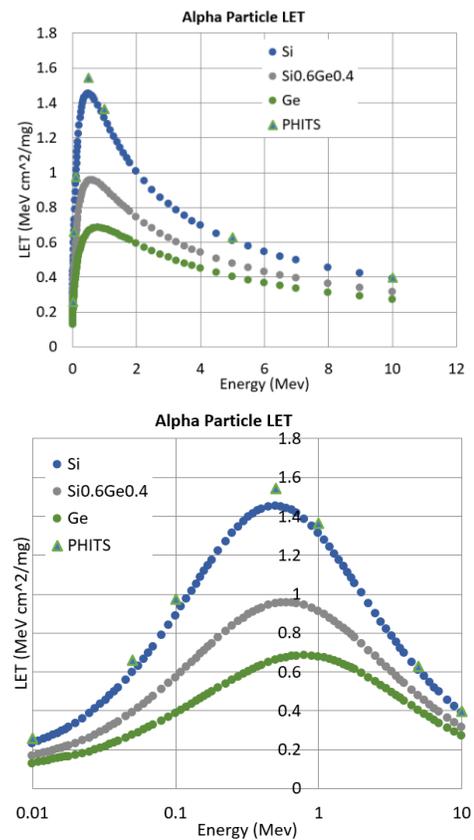

Fig. 2. Alpha particle LET as a function of incident energies in Si, $Si_{0.6}Ge_{0.4}$, and Ge extracted from SRIM simulations. The LET in Si is also extracted using PHITS. Top: Linear scale in x-axis. Bottom: Logarithmic scale in x-axis.

## II. LET EXTRACTION

*Ab initio* tools are used to extract the LET of α−particle and neutron in $Si_{1-x}Ge_x$ (x=0 to 1) at various energies. SRIM is used to extract the LET as a function of α−particle incident energies in $Si_xGe_{1-x}$ (only x=1, 0.6, 0 are shown in Fig. 2). To verify the results, PHITS is also used in the Si case, and it gives similar results. It is found that 0.5MeV α−particle results in the highest LET in Si and the peak shifts gradually to 0.8MeV in pure Germanium.

Since neutron has no charge and it relies on secondary ion generation to ionize electrons in $Si_xGe_{1-x}$, GEANT4, instead of SRIM, is used to calculate its LET. PHITS can be used for this purpose too but was not studied. Fig. 3 shows the methodologies to calculate the LET. The neutron loses energy to create secondary ions after traveling a distance of L1. The secondary ion then loses energy (dE) to ionize electrons in the target. Eventually, the secondary ion may create a tertiary particle due to a collision after traveling a distance of L2. Considering the dimension of the FinFET and stacked-NW being small, only the ionization due to the first secondary ion is considered. The LET can be calculated as dE/(L1+L2) (method 1) or dE/L2 (method 2). Method 1 gives very low LET (about 100 times lower than method 2) because the traveling path of neutron before a collision is long. Therefore, method 2 is used. Ten SRIM simulations were done and the results are averaged and showed in Fig. 4. It can be seen that, unlike α−particle, neutron LET increases with energy monotonically.

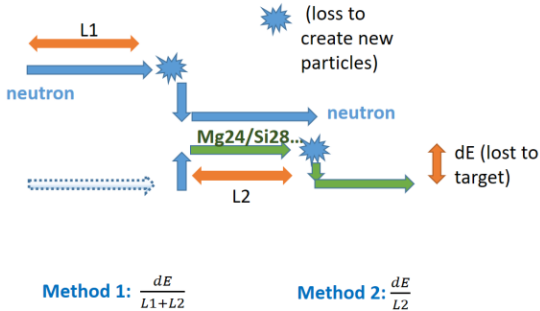

Fig. 3. Methodologies to calcuate the LET of neutron in SRIM. Method 2 is used in this study.

## III. DEVICE SIMULATIONS

TCAD Sentaurus is used in this study. Sentaurus Process is used for structure creation [8] and Sentaurus Device [9] is used for device simulation. N-type and p-type FinFET and nanowires are created through process simulation based on layout. In particular, the stress effect is simulated by taking the anisotropic mechanical properties of Silicon into account. Tensile stress is achieved in n-type devices by creating Si:C source/drain epitaxial while compressive stress in the p-type channel is achieved by creating SiGe pocket in the source/drain. Fig. 5 shows the structures created and their 2D cross-sections. The nanowire stack is created by assuming a similar vertical etching profile/slope as in fin etching. The nanowire stacks are sized such that they give a similar ON-state current as the FinFET. All devices have $L_G$ = 20nm, which corresponds to the 5nm technology node [1].

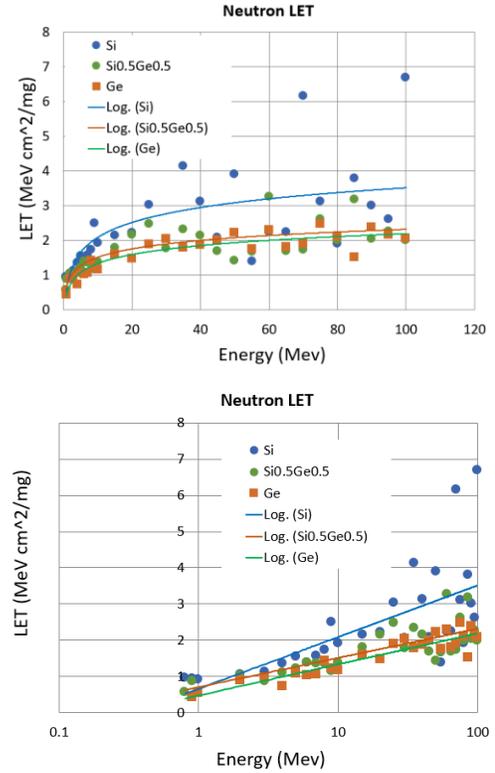

Fig. 4. Neutron LET as a function of incident neutron energy in Si, $Si_{0.5}Ge_{0.5}$ and Ge extracted from GEANT4 simulations. Fitting lines are logathmic. Top: Linear scale in x-axis. Bottom: Logarithmic scale in x-axis.

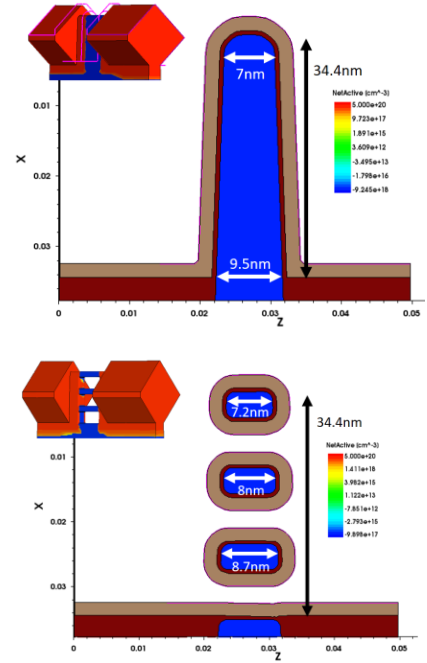

Fig. 5. Cross-sections of n-type FinFET (top) and NW (bottom) used in this study. Both has $L_G$=20nm.

The drift-diffusion model with density gradient to account for quantum effect is used to simulate the device's electrical performance ($I_D V_G$). Based on [1], $V_{DD}$ is set to 0.7V. Due to the small gate length, a low field ballistic mobility model is used [10]. The ballistic mobility model and high field saturation model are calibrated together against Monte Carlo simulation using Sentaurus Device Monte Carlo [11]. The ballistic mobility is given by

$$\mu_{bal} = k L_{ch} \quad (1)$$

where $L_{ch}$ is the channel length and $k$ is the constant where it is found to be 6.4 cm$^2$/Vsnm and 200 cm$^2$/Vsnm for NMOS and PMOS, respectively.

To study the radiation hardness of the transistors, they are biased at off state with $|V_D| = 0.7V$ and $|V_{GS}| = 0V$. Radiation is incident at various locations with LET = 0.0125pC/μm. Note that in the device simulation, the LET is translated directly to the electron-hole pairs to be generated in the particle path. Therefore, to correlate the TCAD LET to the numbers in Fig. 2 and Fig. 4, one needs to take the conversion factor of 0.28 into account (for the fact that not all energies deposited will generate e/h pairs) in addition to unit conversions. Therefore, 0.0125pC/μm in TCAD corresponds to 1.34MeVcm$^2$/mg in Fig. 2 and Fig. 4, which are generated around the peak by α−particle (0.3-0.8MeV) or by 3MeV neutron.

The induced charge of single FinFET and NW devices at various particle strike locations are shown in Fig. 6. It shows that stacked-NW is more robust than FinFET. It is observed that "through S/D" and "top" strikes induce the most charge because the particle traverses the largest active region. The most vulnerable location is position 34, where the particle strikes from the top and traverses through the drain junction vertically where there is a high electric field to prevent e/h pair recombination. It is also worth noting that if bulk FinFET and NW are used (as in this case), they are both vulnerable when the radiation strikes at position 8, where it has the least gate control. Overall, FinFET has about 1.5 to 2 times higher charge generation than NW at vulnerable locations.

## IV. SRAM RADIATION HARDNESS

FinFET and NW SRAM are then constructed using the same process flow with the appropriate mask. Fig. 7 shows the 3D view of the structures. The SRAM is biased at hold state while the bitline (BL) and bitline bar (BLB) have been pre-charged to levels opposite to the stored value in the SRAM cell being studied (Fig. 8). This corresponds to the situation that another cell in the same column is being written. Based on the result in the single device radiation hardness study, four striking locations are selected. Firstly, top strikes at position 34 at off-state NMOS N2 (34N) and PMOS P1 (34P) are simulated as they are expected to be the most vulnerable (Fig. 6). Then, lateral strike at positions 4 and 8 at N2 is also selected because the particles will strike two off-state NMOS (N2 and ACC2) at the same time, due to the layout effect [12]. Various TCAD LET's are simulated and the flipping energies are shown in Table I.

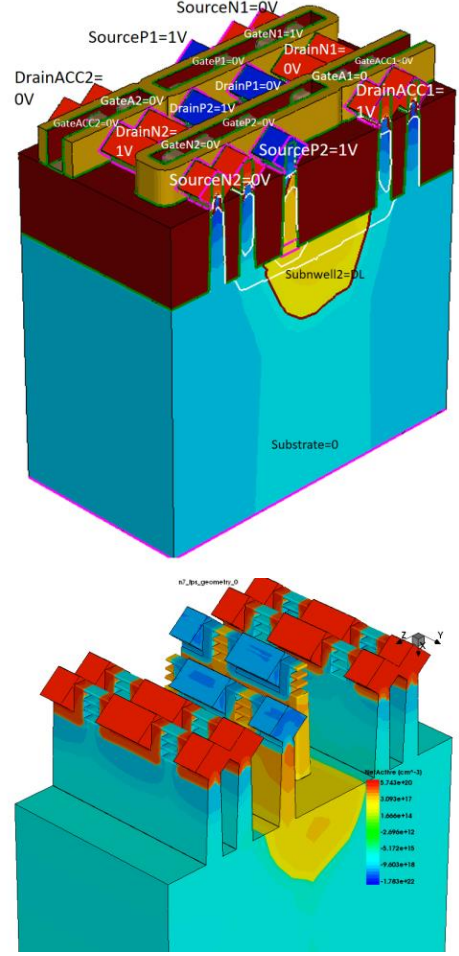

Fig. 7. 3D view of NW SRAM. Top: with labels and all materials. Bottom: only Si is shown.

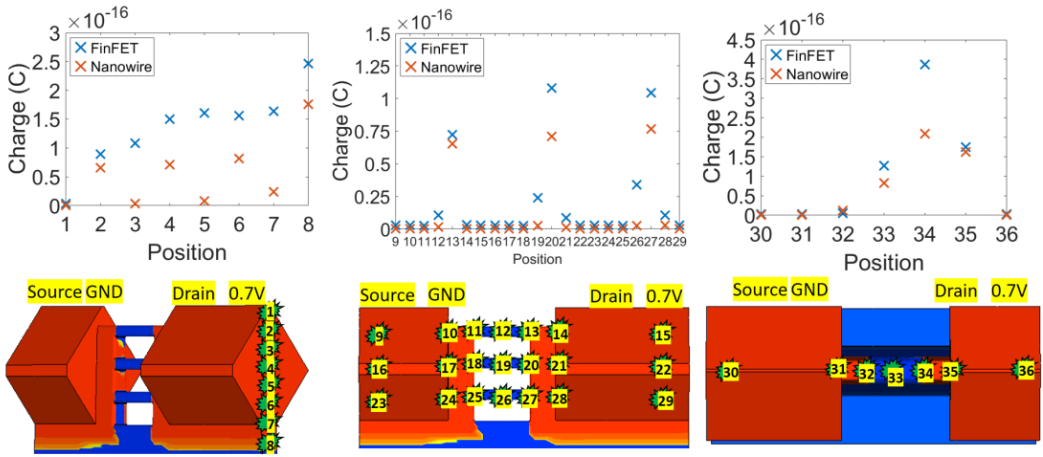

Fig. 6. Top: Charge induced in offstate n-type FinFET and Nanowire when particle with LET = 0.0125pC/μm is incident at various locations (1-36) and different directions (through SD, from the side, and from the top). Bottom: Striking location and biased conditions. Only nanowire is shown.

It can be seen that while the top strike at position 34 is the worst in single device simulation (highest induced charge), "through S/D" strikes is the worst in SRAM (lowest flipping LET). This is because a single particle can turn on two off-state transistors at the same time. As a result, the robust position 4 has similar flipping energy as position 34 in SRAM, while the generated charge is only 37% of that of position 34 in a single device. Therefore, it is very important to perform full layout radiation hardness simulation.

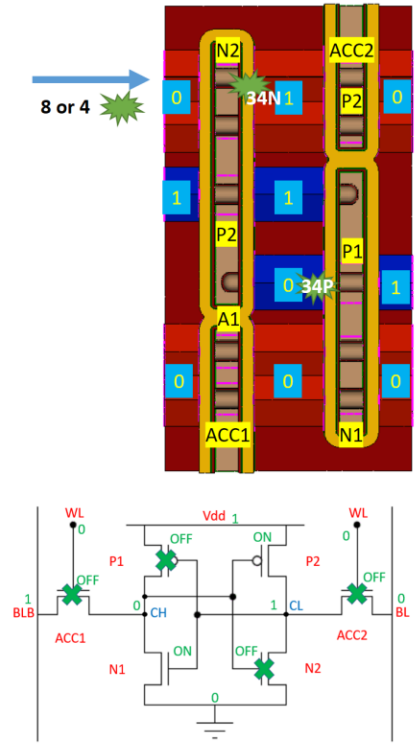

Fig. 8. Top: Striking positions in the SRAM. 34N and 34P are "top strikes" and 8 and 4 are "through SD" strike. Bottom: the corrsponding circuit diagram during strike.

TABLE I. FLIPPING ENERGIES OF SRAMS

| Position | FinFET | | Nanowire | |
|---|---|---|---|---|
| | TCAD LET (pC/mm) | LET (MeVcm²/mg)[a] | TCAD LET (pC/mm) | LET (MeVcm²/mg)[a] |
| 4 (SD Middle) | 0.0035 | 0.38 | 0.0085 | 0.92 |
| 8 (SD Bottom) | 0.0015 | 0.16 | 0.0045 | 0.48 |
| 34N (Top) | 0.003 | 0.32 | 0.01 | 1.08 |
| 34P (Top) | 0.004 | 0.43 | 0.014 | 1.5 |

[a.] Conversion factor of 0.28 used

If the SRAM is designed so that there is no bulk region (e.g SOI SRAM), then the most vulnerable position becomes 34N for FinFET and position 4 for NW, respectively. Based on the LET obtained in Table I, we can convert the flipping energies to α−particle or neutron energies that will flip the SRAM by using Fig. 2 and Fig. 4. They are shown in Table II.

TABLE II. FLIPPING PARTICLE ENERGIES

| Particle Energy (MeV) | α-particle | neutron |
|---|---|---|
| FinFET Bulk | 0.01-10 | >0.44 |
| FinFET non-Bulk | 0.02-10 | >0.57 |
| NW Bulk | 0.038-7 | >0.75 |
| NW non-Bulk | 0.1-2.25 | >1.50 |

It can be seen that NW is the most robust. If non-bulk NW is used, only α−particles with energies between 0.1MeV and 2.25MeV can flip the cell. It is expected that NW SRAM might be immune to α−particles if the design can be improved.

## V. CONCLUSION

Using an *ab initio*-TCAD framework, it is found that FinFET and NW are the most vulnerable in the S/D direction strikes and that 5nm node NW SRAM is more robust than FinFET SRAM. However, the bulk channel in NW can be a bottleneck and should be designed carefully in DTCO. We also project the possibility of designing NW SRAM that is immune to α−particles.

ACKNOWLEDGMENT

This work is supported by DoD (NSWC Crane grant number N00164-19-1-1001).